\documentclass[aps,prd,12pt,preprint,nofootinbib,superscriptaddress,tightenlines,titlepage,floatfix]{revtex4}
 
\usepackage{graphicx}% Include figure files
\usepackage{rotating}
\usepackage{dcolumn}% Align table columns on decimal point
\usepackage{bm}% bold math
\usepackage{longtable}% Include figure files

\begin{document}

\newcommand{\stavg}{(\aDM-\bias)_{\rm s}}
\newcommand{\myfigwid}{6.4in}
\newcommand{\ThreePi}{\pi^+\pi^-\piz}
\newcommand{\etaprime}{\eta\,'}
\newcommand{\pipijpsi}{\pi^+\pi^-\jpsi}
\newcommand{\pizpizjpsi}{\pi^0\pi^0\jpsi}
\newcommand{\kkjpsi}{K^+K^-\jpsi}
\newcommand{\etajpsi}{\eta\jpsi}
\newcommand{\pizjpsi}{\piz\jpsi}
\newcommand{\etaetajpsi}{\eta\eta\jpsi}
\newcommand{\etaprimejpsi}{\etaprime\jpsi}
\newcommand{\threepijpsi}{\ThreePi\jpsi}
\newcommand{\pipipsiprime}{\pi^+\pi^-\psi(2S)}
\newcommand{\etapsiprime}{\eta\psi(2S)}
\newcommand{\pipiphi}{\pi^+\pi^-\phi}
\newcommand{\gammachicone}{\gamma\chi_{c1}}
\newcommand{\gammachictwo}{\gamma\chi_{c2}}
\newcommand{\threepichicone}{\ThreePi\chi_{c1}}
\newcommand{\threepichictwo}{\ThreePi\chi_{c2}}
\newcommand{\omegachiczero}{\omega\chi_{c0}}

\newcommand{\byppj}{{\cal B}(\pipijpsi)}
\newcommand{\bjll}{{\cal B}(J/\psi\to \ell^+\ell^-)}
\newcommand{\brff}{{\cal B}(\psi(4040)\to X\jpsi)}
\newcommand{\brfos}{{\cal B}(\psi(4160)\to X\jpsi)}
\newcommand{\fsb}{$f_{SB}$}
\newcommand{\ie}{{\it i.e.}}
\newcommand{\eg}{{\it e.g.}}
\newcommand{\BABAR}{{\sc BaBar}}
\newcommand{\dimu}{{\mu^+\mu^-}}
\newcommand{\diel}{{e^+e^-}}
\newcommand{\dipi}{{\pi^+\pi^-}}
\newcommand{\dika}{{K^+K^-}}
\newcommand{\fog}{{4\gamma}}
\newcommand{\dipiz}{{\pi^0\pi^0}}
\newcommand{\jpsi}{{J/\psi}}
\newcommand{\chicJ}{\chi_{cJ}}
\newcommand{\psip}{{\psi(2S)}}
\newcommand{\etap}{\eta\,'}
\newcommand{\etal}{{\sl et al.}}
\newcommand{\piz}{{\pi^0}}
\newcommand{\piza}{{\pi^0_a}}
\newcommand{\pizb}{{\pi^0_b}}
\newcommand{\dilep}{{\ell^+\ell^-}}
\newcommand{\qqbar}{q \bar q}
\newcommand{\gev}{\,\mbox{GeV}}
\newcommand{\mev}{\,\mbox{MeV}}
\newcommand{\PiPiJ}{$\pi^+\pi^-\jpsi$}
\newcommand{\BRPiPiJ}{$\pi^+\pi^-\jpsi$ (tight)}
\newcommand{\PizPizJ}{$\pi^0\pi^0\jpsi$}
\newcommand{\PizJ}{$\pi^0\jpsi$}
\newcommand{\EtaJ}{$\eta\jpsi$}
\newcommand{\EtaJGG}{$\eta(\to \gamma\gamma)\jpsi$}
\newcommand{\EtaJThreePi}{$\eta(\to \pi^+\pi^-\pi^0)\jpsi$}
\newcommand{\XJ}{$X\jpsi$}
\newcommand{\OnePiJ}{$\pi^\pm(\pi^\mp)\jpsi$}
\newcommand{\OnePizJ}{$\pi^0(\pi^0)\jpsi$}
\newcommand{\ChicZero}{$\gamma \chi_{c0} \to \gamma\gamma\jpsi$}
\newcommand{\ChicOne}{$\gamma \chi_{c1} \to \gamma\gamma\jpsi$}
\newcommand{\ChicTwo}{$\gamma \chi_{c2} \to \gamma\gamma\jpsi$}
\newcommand{\PiPiJRhoPi}{$\pi^+\pi^-\jpsi,\ \jpsi \to \rho\pi $}
\newcommand{\PiPiJPiPi}{$\pi^+\pi^-\jpsi,\ \jpsi \to \pi\pi $}

\newcommand{\PiPiJMuMu}{$\pi^+\pi^-\mu^+\mu^-$}
\newcommand{\PiPiJEE}{$\pi^+\pi^-e^+e^-$} 
\newcommand{\PizPizJMuMu}{$\pi^0\pi^0\mu^+\mu^-$}
\newcommand{\PizPizJEE}{$\pi^0\pi^0e^+e^-$}
\newcommand{\PizJMuMu}{$\pi^0\mu^+\mu^-$}
\newcommand{\PizJEE}{$\pi^0e^+e^-$}
\newcommand{\EtaJMuMu}{$\eta\mu^+\mu^-$}
\newcommand{\EtaJEE}{$\eta e^+e^-$}
\newcommand{\EtaJMuMuGG}{$\eta(\to\gamma\gamma)\mu^+\mu^-$}
\newcommand{\EtaJEEGG}{$\eta(\to\gamma\gamma)e^+e^-$}
\newcommand{\EtaJMuMuThreePi}{$\eta(\to\pi^+\pi^-\pi^0)\mu^+\mu^-$}
\newcommand{\EtaJEEThreePi}{$\eta(\to\pi^+\pi^-\pi^0)e^+e^-$}
\newcommand{\XJMuMu}{$X\mu^+\mu^-$}
\newcommand{\XJEE}{$Xe^+e^-$}
\newcommand{\OnePiJEE}{$\pi^\pm(\pi^\mp)e^+e^-$}
\newcommand{\OnePiJMuMu}{$\pi^\pm(\pi^\mp)\mu^+\mu^-$}
\newcommand{\OnePizJEE}{$\pi^0(\pi^0)e^+e^-$}
\newcommand{\OnePizJMuMu}{$\pi^0(\pi^0)\mu^+\mu^-$}
\newcommand{\ChicZeroMuMu}{\ChicZero $\jmumu$}
\newcommand{\ChicZeroEE}{\ChicZero $\jee$}
\newcommand{\ChicOneMuMu}{\ChicOne $\jmumu$}
\newcommand{\ChicOneEE}{\ChicOne $\jee$}
\newcommand{\ChicTwoMuMu}{\ChicTwo $\jmumu$}
\newcommand{\ChicTwoEE}{\ChicTwo $\jee$}

\newcommand{\BRPiPiJMuMu}{$\pi^+\pi^-\mu^+\mu^-$ (tight)}
\newcommand{\BRPiPiJEE}{$\pi^+\pi^-e^+e^-$ (tight)} 

\newcommand{\ChicZeroJMuMu}{$\gamma\gamma\mu^+\mu^-(\chi_{c0})$}
\newcommand{\ChicZeroJEE}{$\gamma\gamma e^+e^-(\chi_{c0})$}
\newcommand{\ChicOneJMuMu}{$\gamma\gamma\mu^+\mu^-(\chi_{c1})$}
\newcommand{\ChicOneJEE}{$\gamma\gamma e^+e^-(\chi_{c1})$}
\newcommand{\ChicTwoJMuMu}{$\gamma\gamma\mu^+\mu^-(\chi_{c2})$}
\newcommand{\ChicTwoJEE}{$\gamma\gamma e^+e^-(\chi_{c2})$}

\newcommand{\bxg}{\byppj\times\Gamma_{\rm ee}}

\newcommand{\SB}{side band}
\newcommand{\XF}{cross feed}
\newcommand{\ipb}{pb$^{-1}$}
\newcommand{\DM}{\delta}
\newcommand{\aDM}{\langle\DM\rangle}
\newcommand{\aDMi}{\langle\DM_i\rangle}
\newcommand{\aDMm}{\langle\DM_\dimu\rangle}
\newcommand{\aDMp}{\langle\DM_\dipi\rangle}
\newcommand{\bias}{\beta}
\newcommand{\sccont}{continuum subtraction}

\newcommand{\gga}{\gamma\gamma}
\newcommand{\tpi}{\pi^+\pi^-\pi^0}
\newcommand{\tpz}{3\pi^0}
\newcommand{\ppg}{\pi^+\pi^-\gamma}
\newcommand{\eeg}{e^+e^-\gamma}
\newcommand{\meta}{M_{\eta}}
\newcommand{\metapdg}{\meta^{\rm PDG}}
\newcommand{\metamc}{\meta^{\rm MC}}
\newcommand{\inv}{invisible}

\newcommand{\brgg}{{\cal B}(\eta\to\gamma\gamma)}
\newcommand{\brtpi}{{\cal B}(\eta\to\pi^+\pi^-\pi^0)}
\newcommand{\brtpz}{{\cal B}(\eta\to3\pi^0)}
\newcommand{\brppg}{{\cal B}(\eta\to\pi^+\pi^-\gamma)}
\newcommand{\breeg}{{\cal B}(\eta\toe^+e^-\gamma)}
\newcommand{\brinv}{{\cal B}(\eta\toinvisible)}
\newcommand{\bretaj}{{\cal B}(\psip\to\eta\jpsi)}
\newcommand{\brpizj}{{\cal B}(\psip\to\pi^0\jpsi)}
\newcommand{\brtpij}{{\cal B}(\psip\to\pi^+\pi^-\jpsi)}
\newcommand{\brtpzj}{{\cal B}(\psip\to\pi^0\pi^0\jpsi)}
\newcommand{\brggj}{{\cal B}(\psip\to\gga\jpsi)}

\newcommand{\Mpppdg}{M^{\rm PDG}_\psip}
\newcommand{\Mjppdg}{M^{\rm PDG}_\jpsi}
\newcommand{\Mpizpdg}{M^{\rm PDG}_\piz}

\preprint{CLNS~07/2003}       % for CLNS notes
\preprint{CLEO~07-09}         % for CLNS notes

\title{\Large 
Measurement of the $\eta$-Meson Mass using
$\psip\to\eta\jpsi$}

\author{D.~H.~Miller}
\author{B.~Sanghi}
\author{I.~P.~J.~Shipsey}
\author{B.~Xin}
\affiliation{Purdue University, West Lafayette, Indiana 47907, USA}
\author{G.~S.~Adams}
\author{M.~Anderson}
\author{J.~P.~Cummings}
\author{I.~Danko}
\author{J.~Y.~Ge}
\author{D.~Hu}
\author{B.~Moziak}
\author{J.~Napolitano}
\affiliation{Rensselaer Polytechnic Institute, Troy, New York 12180, USA}
% ...
%\author{(CLEO Collaboration)} %FOR PRD_SPECIAL_CHANGEME
\collaboration{CLEO Collaboration} %FOR PRL,CLNS
\noaffiliation

\date{July 11, 2007}

\begin{abstract}
We measure the mass of the $\eta$ meson 
using $\psip$$\to$$\eta\jpsi$ events acquired with the CLEO-c detector
operating at the CESR $e^+e^-$ collider.
Using the four decay modes $\eta$$\to$$\gga$, $\tpz$, $\tpi$, and $\ppg$,
we find $\meta$=547.785$\pm$0.017$\pm$0.057~MeV, in which
the first uncertainty is statistical and the second systematic. 
This result has an uncertainty comparable to the two most precise
previous measurements and is consistent with that of NA48,
but is inconsistent at the level of 6.5$\sigma$  
with the much smaller mass obtained by GEM.

\end{abstract}

%\pacs{14.40.Aq}

\maketitle

  The $\eta$ meson, the second-lightest pseudoscalar,
is commonly understood as being predominantly in the SU(3)-flavor octet
with a small singlet admixture, so that
it has comparable $u\bar{u}$, $d\bar{d}$, and $s\bar{s}$ 
content and virtually no gluonium component~\cite{rosner,escrib}.
Its mass is of fundamental importance to understanding
the octet-singlet mixing as well as the gluonium content
of both $\eta$ and $\eta^\prime$~\cite{michael}, although theoretical
and phenomenological precision on related predictions~\cite{gerard,kekez} 
has not yet matched that of experiment.
On the experimental side, there has long been
a situation of conflicting $\meta$ measurements
that improvements in precision have been unable to resolve.
Indeed, the current status is the worst it
has ever been, with a confidence level (C.L.) of
0.1\% that the measurements are consistent~\cite{PDG2006}.
In 2002 the world average was
$\meta$=547.30$\pm$0.12~MeV~\cite{PDG2002} and
results included in it were generally consistent with one another,
but only because 1960's-era
bubble chamber experiments, all of which favored a larger
$\meta$, had been dropped~\cite{PDG1998}. 
Then NA48,
based on exclusively reconstructed $\eta$$\to$3$\piz$ decays,
reported
$\meta$=547.843$\pm$0.030$\pm$0.041~MeV~\cite{NA48},
which appeared to vindicate the dropped
experiments. In 2005 GEM reported
$\meta$=547.311$\pm$0.028$\pm$0.032~MeV~\cite{GEM}
using the $^3$He recoil mass in $p$+$d$$\to$$^3$He+$X$.
The GEM result was consistent with less precise results 
made during the period 1974-1995,
but also was eight standard deviations below that of NA48.
More measurements with sub-100~keV precision
are needed to clarify the matter.

  This work presents a new measurement of $\meta$
using $\psip$$\to$$\eta\jpsi$. Events were acquired with
the CLEO-c detector at the CESR (Cornell Electron %-positron
Storage Ring) symmetric $e^+e^-$ collider. The data sample
corresponds to $\sim$27 million produced 
$\psip$ mesons, of which about  0.8$\times$10$^6$
decay to $\eta\jpsi$. 
  We measure the mass by exploiting
kinematic constraints in the
decay chain $\psip$$\to$$\eta\jpsi$, $\jpsi$$\to$$\dilep$
($\ell^\pm$$\equiv$$e^\pm$ or $\mu^\pm$),
and $\eta$$\to$$\gga$, $\tpz$, $\tpi$, or $\ppg$.
Because both $\psip$ and $\jpsi$ are very narrow
resonances with precisely known masses,
the constraints enable
a significant improvement in $\eta$ mass
resolution over that achieved by the
detector alone. This  is the first $M_\eta$ measurement 
to use $\psip$$\to$$\eta\jpsi$.

The CLEO-c detector is described in detail elsewhere~\cite{CLEO};
it offers 93\% solid angle coverage of precision 
charged particle tracking and an electromagnetic calorimeter
comprised of CsI(Tl) crystals.
The tracking system enables momentum measurements
for particles with momentum
transverse to the beam exceeding 50~MeV/$c$,
and achieves resolution $\sigma_p/p\simeq$0.6\% at $p$=1~GeV/$c$.
The barrel calorimeter reliably measures photon showers down
to $E_\gamma$=30~MeV and has a resolution of
$\sigma_E/E\simeq$5\% at 100~MeV and 2.2\% at 1~GeV.

  Event selection begins with that described
in Ref.~\cite{breta} for the four
$\eta$ decay modes used here. 
We also accumulate samples of $\psip$$\to$$ \dipi\jpsi$, $\dipiz\jpsi$, 
and $\piz\jpsi$ for studies of systematic uncertainties. Every
particle in the decay chain is sought,
and events are separated into those with
$\jpsi$$\to$$\dimu$ and $\jpsi$$\to$$\diel$. 
Leptons are loosely identified and
restricted to $|\cos\theta_{\ell^\pm}|$$<$0.83, where
$\theta$ is the angle of the track with
respect to the incoming positron beam.
Lepton momenta are augmented with calorimeter showers found
within a 100~mrad cone of the initial track direction,
under the assumption that they are produced by bremsstrahlung.
All photon candidates are required to
be located in the central portion of the barrel calorimeter
where the amount of material traversed is smallest
and therefore energy resolution is best:
$|\cos\theta_\gamma|$$<$0.75.
Backgrounds of 1-4\% are present
in each $\eta\jpsi$ sub-sample~\cite{breta}, consisting of cross-feed
from other $\eta$ decays as well as 
other $\psip\to X\jpsi$ decays: $\dipi\jpsi$, $\dipiz\jpsi$,
and $\gamma\chicJ$, $\chicJ$$\to$$\gamma\jpsi$.

  Kinematic constraints are applied in two two-step fits: 
first, the lepton tracks are constrained to a common origination
point (vertex) and thence to the $\jpsi$ mass, 
$\Mjppdg$~\cite{PDG2006};
second, the constrained $\jpsi$, the beam spot 
and the $\eta$ decay products are
constrained to a common vertex and then to the $\psip$ mass, 
$\Mpppdg$~\cite{PDG2006}.
Separate fit quality restrictions are
applied to vertex ($\chi^2_{\rm v}$) and mass ($\chi^2_{\rm m}$) 
constraints.
For $\eta$$\to$$\tpi$, the $\piz$$\to$$\gga$ candidate
is constrained to the $\piz$ mass prior to
the fits described above. The decay $\eta$$\to$$\tpz$
is treated as $\eta$$\to$$ 6\gamma$ because
reliably making a unique set of correct photon-$\piz$ assignments
is not possible; typically several such
assignments per event of comparable probability
exist and are indistinguishable.
To ensure that only the best
measured events survive into the
final sample, the $\chi^2$ restrictions
from Ref.~\cite{breta} are tightened
to $\chi^2_{\rm v}/$d.o.f.$<$10 and  $\chi^2_{\rm m}/$d.o.f.$<$5
for both the $\jpsi$ and $\psip$ constrained fits.

  Alternative event topologies
are used to compare measurements of the $\piz$ mass
to its established value,  
$\Mpizpdg$=134.9766$\pm$0.0006~MeV~\cite{PDG2006},
for two different $\piz$ momentum ranges.
The first ($\piza$) is $\psip$$\to$$\piz\jpsi$,  $\piza$$\to$$\gga$,
which features a monochromatic $\piz$ with $p$\,$\simeq$\,500~MeV/$c$,
and the second ($\pizb$) $\psip$$\to$$\eta\jpsi$, $\eta$$\to$$\tpi$,
$\pizb$$\to$$\gga$, which contains $\piz$'s with $p$\,$\simeq$\,0-250~MeV/$c$.
For these tests, the individual photons (instead of a constrained $\piz$) 
are used in the $M_\psip$ constraint on all final state four-momenta.

\begin{figure}[thp]
\includegraphics*[width=\myfigwid]{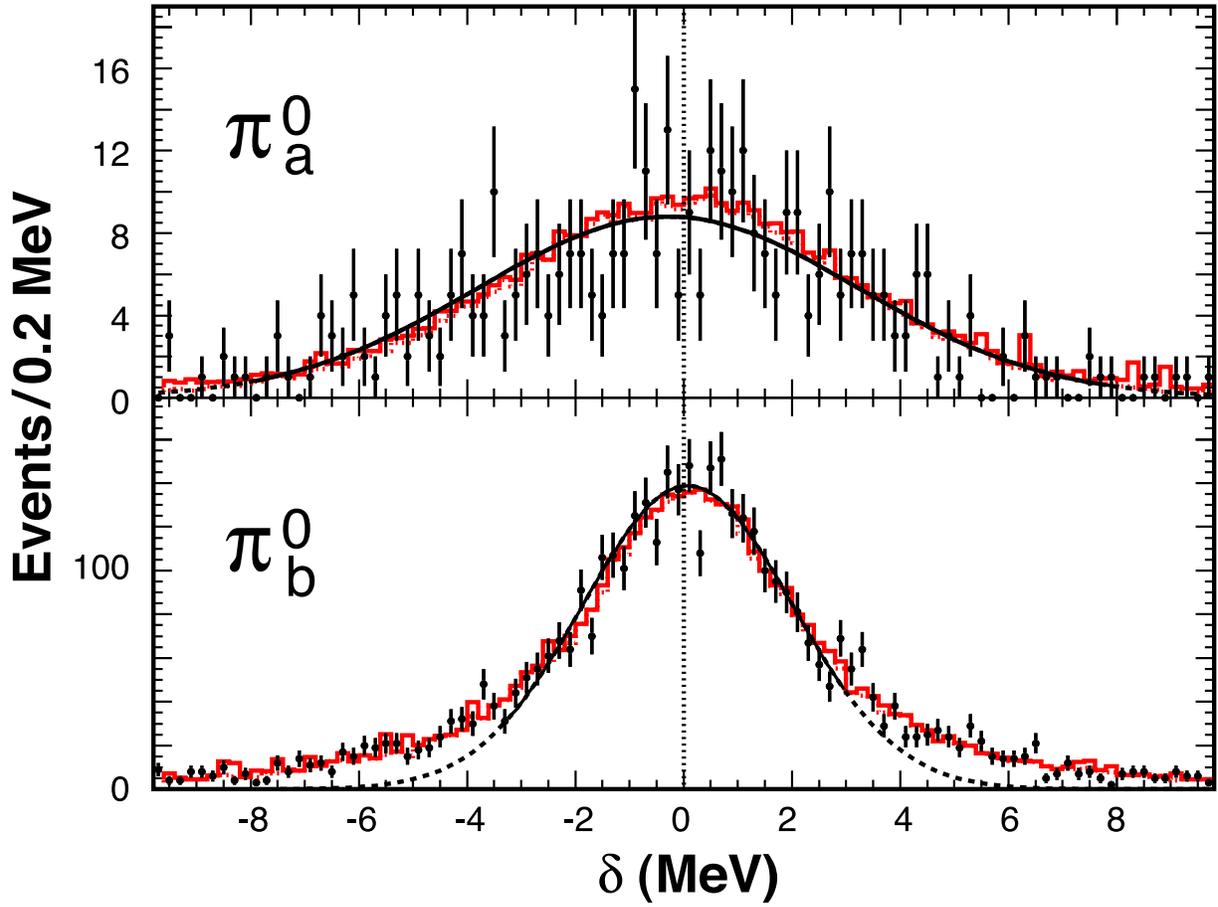}
\caption{Distributions of $\DM$ for two evaluations
of $M_\piz$ using $\piz$$\to$$\gga$ decay,
with the data represented by the points with error bars
and the Gaussian fit overlaid. 
The solid line
portion of the fit indicates the window used for the fit
and the dashed portions its extension.
The solid line histogram represents
MC simulation of signal and backgrounds with 
$M_\piz^{\rm MC}$=$\Mpizpdg$
and $\metamc$=547.78~MeV. 
\label{fig:mpiz} }
\end{figure}

\begin{figure}[thp]
\includegraphics*[width=6.2in]{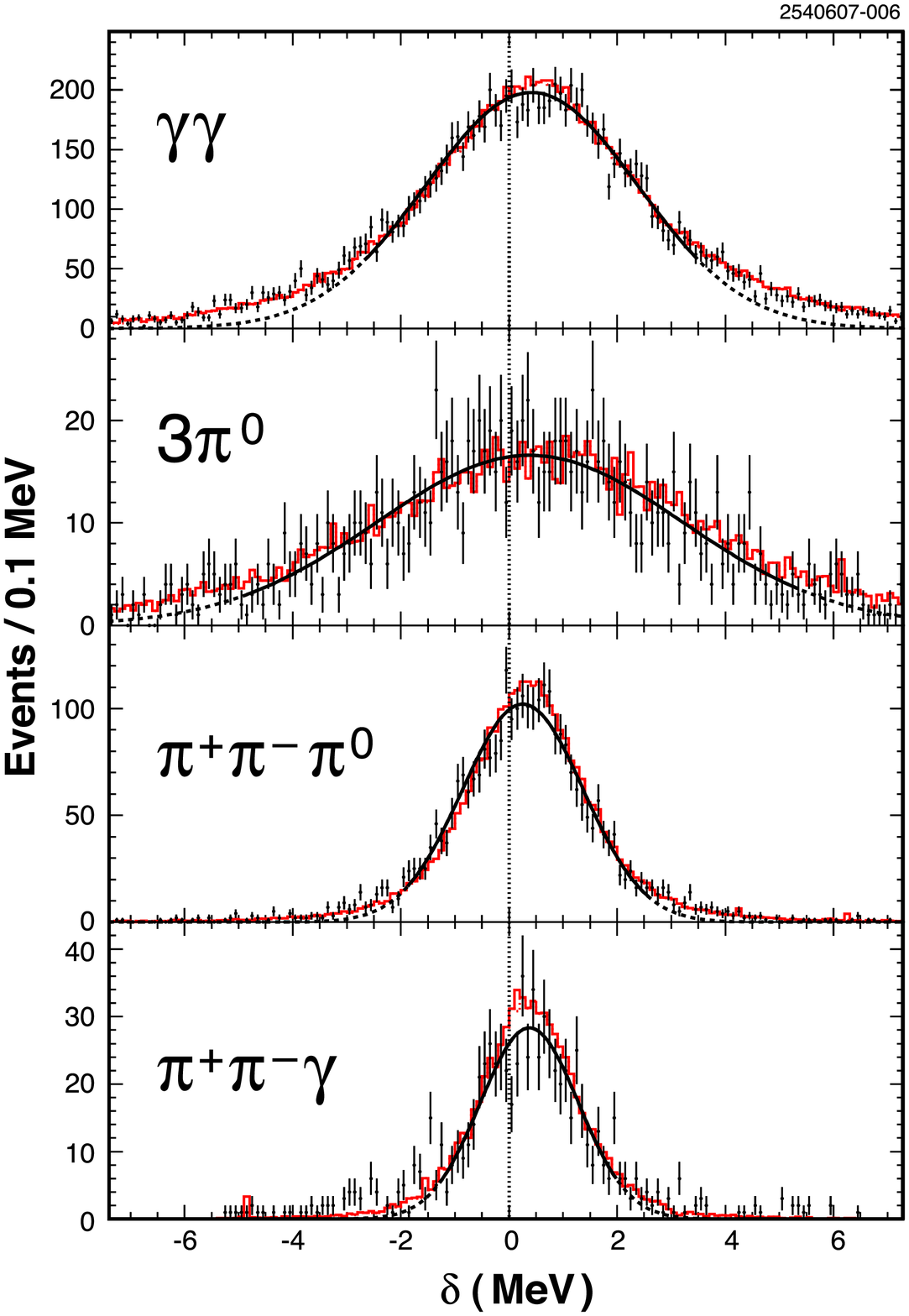}
\caption{Distributions of $\DM$ for $\eta$ decay
modes as shown. Symbols are as defined in Fig.~1. 
\label{fig:meta} }
\end{figure}

  Each event yields an invariant mass $M$ of the 
kinematically-constrained decay products;
a single mass value 
is extracted for each decay mode $i$
by fitting a Gaussian shape to the distribution of
$\DM_i$$\equiv$$M_i$$-$$M_0$,
where $M_0$ is simply a reference value, either the
current Particle Data Group world-average 
$\metapdg$=547.51$\pm$0.18~MeV~\cite{PDG2006},
or, in the case of the $\piz$ cross-check modes, 
$\Mpizpdg$ as given above.
The fits are restricted to
the central portion of each $\DM$ distribution
because the tails outside this region are not represented well by
a single Gaussian form. The fits
span $\pm$1.6$\sigma$ to $\pm$2.0$\sigma$
about the peak $\aDM$, where $\sigma$ is the
fitted Gaussian width, and in all cases
the resulting fit has a C.L. exceeding 1\%.
The distributions of $\DM_i$ for $\piz$ and $\eta$ decay 
with fits are shown
in Figs.~\ref{fig:mpiz} and \ref{fig:meta}, respectively.
Other shapes that might fit the tails,
such as a double Gaussian, have been found to
yield unstable fits and/or do not improve precision
of finding the peak.

   There is an unavoidable 
low-side tail in any monochromatic photon
energy distribution from the CLEO calorimeter. 
It originates from losses sustained 
in interactions prior to impinging
upon the calorimeter and from leakage
outside those crystals used
in the shower reconstruction.
This asymmetric photon energy resolution
function also results in
a small but significant systematic bias
in $\aDM$: for simplicity
of the kinematic fitting formalism,
input uncertainties  are
assumed to be symmetric,
and a bias occurs if they are not. 
This bias in fitted Gaussian mean
is mode-dependent because each
presents a different mix of charged and neutral
particles.

The biases $\bias_i$ are estimated by following
the above-described procedure on MC signal samples.
Each $\bias_i$ is the difference between
the Gaussian peak value of the $\meta$ distribution
and the input $\metamc$. We define the bias as
$\bias_i$$\equiv$$\aDMi_{\rm MC}$,
in which we use
the MC input $\metamc$ instead of $\metapdg$ for $M_0$.
A non-zero value of $\bias_i$ means that, for mode $i$,
the Gaussian peak mass $\aDMi$ is offset
from the true mass and must be corrected.
We evaluate the four $\bias_i$ 
for trial values of $\metamc$
(547.0, 547.3, 547.8, and 548.2~MeV)
that cover the spread of previous measurements.
The biases extracted for these $\metamc$ inputs are consistent,
and the final bias for each mode, shown in Table~\ref{tab:tableres},
is taken as their average.
Bias values for the
$\piza$ and $\pizb$ cross-checks are determined similarly.

  Table~\ref{tab:tableres} summarizes results by decay mode.
Both $\piza$ and $\pizb$ mass values are consistent
with expectations within their respective statistical
uncertainties. 
The total number of reconstructed events
involved in the determination of $\meta$ is 16325.
The four values of $\aDMi-\bias_i$ have an
average, weighted by statistical errors only,
of $\stavg$=277$\pm$17~keV with a $\chi^2$=4.8
for three degrees of freedom (C.L.$\sim$20\%). 

  Systematic errors are summarized in Table~\ref{tab:tablesys}.
  Uncertainties that are uncorrelated mode-to-mode,
including statistical,
are used to determine the weights 
($w_i$=0.19, 0.03, 0.60, and 0.18
for $\gga$, $\tpz$, $\tpi$, and $\ppg$, respectively) 
applied to combine
values from the four modes into the weighted sum 
$(\aDM$$-$$\bias)_{\rm w}$=$\sum_{i=1}^{4}$$w_i$$\times$$(\aDMi$$-$$\bias_i)$=272$\pm$17~keV.

  As the mass distributions are not
perfectly Gaussian, there is some systematic
variation of the peak value with the choice
of mass limits for each fit. However,
as long as the limits
chosen do not result in a confidence
level below 1\% and remain
roughly symmetric about the peak,
such variation is observed to be bounded by approximately
half of a statistical standard deviation.
Hence this value was assigned as a 
conservative estimate of the systematic
uncertainty attributable to the fit limits.

Uncertainties attributable
to imprecision in the masses of the $\jpsi$
(11~keV) and $\psip$ (34~keV) mesons~\cite{PDG2006}
are directly calculated by repeating the analysis 
using an altered $\psip$ or $\jpsi$ mass 
and the deviation in $\aDM$ per ``1$\sigma$'' change from nominal 
taken as the error.

  The bias $\bias_i$ from kinematic fitting 
can be attributed to the effect of
the asymmetric resolution function of photons;
the bias for the four modes varies by
an order of magnitude, with larger values 
corresponding to modes with more photons.
The $\pizb$ cross-check indicates, 
within its 47~keV statistical
precision on $\aDM$$-$$\bias$, that the bias is indeed accurately
estimated. A more sensitive cross-check comes from
comparing the $\meta$ shifts in data and MC
for $\eta\to\tpi$ when
the $\piz$ mass constraint is removed: the MC predicts
an increase in bias of 76$\pm$6~keV compared
to an observed shift in the data of 55$\pm$24~keV
($i.e.$~the shift in bias of 76~keV is verified in the data within
the 24~keV statistical error, which amounts to about a third of MC shift itself).
Based on these comparisons and the generally favorable
agreement~\cite{breta} between data and MC characteristics,
we take one third of the bias central value ($\bias_i$/3) as 
our estimate of the systematic uncertainty.

  By performing mass fits on MC signal
samples with and without simulated backgrounds,
it is determined that for mode $i$, modeled backgrounds 
reduce the bias by the amount
$B_i$: 21$\pm$19, $-$3$\pm$67, 2$\pm$17, and $-$13$\pm$27 keV
for the $\gga$, $\tpz$, $\tpi$, and $\ppg$ channels,
respectively, where uncertainties listed are statistical. 
The unmodeled background in the $\ppg$ sample~\cite{breta}
is estimated to have an effect on $\meta$ that is negligible
compared to the 27~keV uncertainty on the modeled background 
for this mode. 
After weights are applied,
the net effect is a positive offset
to $\meta$, of $B_{\rm w}$=3$\pm$12~keV.

Uncertainties in charged particle
momentum and calorimeter energy scale
are evaluated by shifting
those scales by the appropriate amount
and repeating the analysis.
The charged particle momentum scale
is confirmed at high momentum
($\sim$1.5~GeV/$c$) using unconstrained 
$\jpsi$$\to$$\dimu$ decays from
$\psip$$\to$$\dipi\jpsi$ and
$\dipiz\jpsi$ events.
A low-momentum (75-500~MeV/$c$) calibration,
which is more relevant to our measurement 
of $M_\eta$, can be made by comparing the 
mass of $\psip$$\to$$\dipi\jpsi$ to $\Mpppdg$ 
with no kinematic constraints
on the $\dipi$ but
with a mass-constrained $\jpsi$$\to$$\dilep$;
this checks the $\dipi$ momentum scale
because the accurately known $\jpsi$ mass
takes up $\sim$84\% of the available energy.
Events of both types are 
subjected to Gaussian fits 
to mass difference variables similar to
$\DM$, in the first case to
$\DM_\dimu$$\equiv$$M(\dimu)$$-$$\Mjppdg$,
and in the second to
$\DM_{\dipi}$$\equiv$$M(\dipi\jpsi)$$-$$\Mpppdg$.
For MC simulation, where
we can employ our perfect knowledge of
the magnetic field for the
momentum scale,
the means and statistical errors of these measures
are $\aDMm$=$-$90$\pm$22~keV
and $\aDMp$=2$\pm$3~keV.
This demonstrates that this technique
is accurate to $|\aDMm|/M_\jpsi$$\simeq$3$\times$10$^{-5}$
and $|\aDMp|/(M_\psip-M_\jpsi)$$\simeq$1$\times$10$^{-5}$.
The magnetic field scale in data is tuned
to that value which keeps both of these
means close to zero: with this setting, measurements
yield $\aDMm$=7$\pm$46~keV 
and $\aDMp$=$-$23$\pm$6~keV,
indicating a similar level of sensitivity as
the MC samples. Therefore
we quote a relative momentum scale
accuracy of 0.01\% and use this value
for our 1$\sigma$ systematic variation.

  Several processes are used for the
calorimeter calibration: inclusive 
$\piz$ decays to $\gga$ (where we can
constrain $M(\gga)$ to a known mass),
$\diel\to\dilep\gamma$ (in which energy-momentum
conservation and well-measured track momenta allow
constraint of the photon energy), and $\psip$$\to$$\gamma\chicJ$
(where the transition photon energies are known well).
The photon energy scale is calibrated to
give the correct peak, \ie\ the most probable value, for any 
monochromatic photon energy distribution.
These calibrations are combined and
result in an overall energy scale
known to 0.6\% or better over the energy range
(30-400~MeV) relevant for photons from the slow
$\eta$-mesons produced in $\psip$$\to$$\eta\jpsi$.

  Any deviation from ideal in momentum
or energy scale is substantially
damped by the mass constraints, as is
evident from Table~\ref{tab:tablesys}:
 the relative momentum (energy) scale
uncertainty of 0.01\% (0.6\%)
induces, at most, $\sim$1 ($\sim$5) parts in 10$^5$
shift in $\eta$-mass scale.

  We have also computed $\stavg$
when the decays occur in combination with either 
$\jpsi$$\to$$\dimu$ or $\jpsi$$\to$$\diel$
separately; its value for $\jpsi$$\to$$\diel$ events
is higher than that for $\jpsi$$\to$$\dimu$ by 98$\pm$34~keV (+2.9$\sigma$),
where the error is statistical only.
Broken down by mode, this difference 
is  71$\pm$57~keV (+1.2$\sigma$) for $\gga$, 
$-$82$\pm$208~keV ($-$0.4$\sigma$) for $\tpz$,
119$\pm$49~keV (+2.4$\sigma$) for $\tpi$, 
and 141$\pm$91~keV (+1.5$\sigma$) for $\ppg$.
We do not observe such an effect in MC simulations.
Further investigations, detailed below,
reveal no firm explanation.
To allow for a hidden systematic effect,
we add an MC modeling uncertainty of 46~keV;  
it is the dominant uncertainty in this analysis.

In order to investigate the possibility that 
there could be an unmodeled systematic pull of
$\jpsi$$\to$$\diel$ or $\jpsi$$\to$$\dimu$
decays in the kinematic fitting process, we examine
$\psip$$\to$$\dipi\jpsi$ decays.
If such an effect existed, 
we would expect to observe a shift between the dipion mass
computed with the {\sl constrained} pion ($\pi^\pm_c$) momenta relative to
the {\sl unconstrained} ($\pi^\pm_u$) values.
However, we find no evidence for such a pull: defining
$f_\dilep$$\equiv$$\langle
M(\pi^+_c\pi^-_c)$$-$$M(\pi^+_u\pi^-_u)\rangle_\dilep$, 
the difference $f_\diel - f_\dimu$=13$\pm$11~keV
in data and 7$\pm$9~keV in MC simulations
(errors shown are statistical).

  To investigate the effect of less-well-measured
events upon the analysis in general and the
$\diel$-$\dimu$ discrepancy in particular,
we have repeated the analysis after tightening
the kinematic fitting restrictions from
$\chi^2_{\rm v}/$d.o.f.$<$10 and  $\chi^2_{\rm m}/$d.o.f.$<$5 
on $\jpsi$ and $\psip$ constrained fits
to 5 and 2, respectively,
losing about 40\% of the original events. 
The net offset from
backgrounds changes from 3$\pm$12~keV to $-$8$\pm$14~keV. 
The overall
final $\eta$ mass, including the background offset,
changes by $-$16$\pm$17~keV, demonstrating
stability of the measured mass with respect to
the kinematic fit quality.
For this restrictive selection,
the $\meta$ difference between $\diel$ and $\dimu$
events in data goes {\sl down} to
48$\pm$44~keV, whereas the MC difference remains near zero. 
It appears that the $\diel$-$\dimu$ discrepancy
moderates for this class of events, but
the statistical precision is not conclusive.

In order to study dependence of $\meta$ upon 
the time of data collection, 
we divide the data into nine contiguous and
consecutive data-taking periods.
One mass from each period is obtained by
averaging the results obtained from the four
modes with statistical weights.
The $\chi^2$ for the nine values to be consistent 
with their statistically-weighted average is 9.5 for 8 degrees
of freedom (C.L.=25\%), demonstrating the absence of
any time-dependent systematics.

  After combining the $(\aDMi$$-$$\bias_i)$ values in 
Table~\ref{tab:tableres} using the quoted weights, including 
the aforementioned net effect of backgrounds $B_{\rm w}$, and 
adding the $\metapdg$ offset, our result is
$\meta$=($\aDM$$-$$\bias$)$_{\rm w}$+$B_{\rm w}$+$\metapdg$= 
547.785$\pm$0.017$\pm$0.057~MeV,
where the first error is statistical and the second systematic. 
This result has comparable precision to both NA48 and
GEM measurements, but is consistent with the former and 6.5 
standard deviations larger than the latter. All four prominent 
$\eta$ decay modes contribute to this result, and each independently 
verifies a significantly larger $\meta$ than obtained by GEM
($\sim$2.0$\sigma$ in $\tpz$, more for each of the other three).

We gratefully acknowledge the effort of the CESR staff
in providing us with excellent luminosity and running conditions.
This work was supported by
the A.P.~Sloan Foundation,
the National Science Foundation,
the U.S. Department of Energy, and
the Natural Sciences and Engineering Research Council of Canada.

\clearpage

\begin{table}[t]
\setlength{\tabcolsep}{0.25pc}
\catcode`?=\active \def?{\kern\digitwidth}
\caption{For each $\piz$ or $\eta$ decay mode,
the number of events $N$, the Gaussian width on the mass
distribution of those data events,
$\sigma$, the values of $\DM$, $\bias$ (from MC),
and the difference $\aDM$$-$$\bias$,
(see text).
Uncertainties shown are statistical.
}
\label{tab:tableres}
\begin{center}
\begin{tabular}{lrrrrrr}
\hline
\hline
Channel & $N$\ \ & $\sigma$\ \ \ & $\aDM$\ \ \ \ & $\bias$\ \ \ \ & $\aDM$$-$$\bias$\ \  \\
        &     &   (MeV)  &  (keV)\ \ \ & (keV)\ \  & (keV)\ \  \\
\hline
$\gga$ ($\piza$) &  420 & 3.51 & $-$285$\pm$195 & $-$51$\pm$21 & $-$234$\pm$196 \\
$\gga$ ($\pizb$) & 4692 & 1.94 & 74$\pm$~\,46 & 128$\pm$~\,8 & $-$54$\pm$~\,47 \\
$\gga$           &11140 & 1.96 & 419$\pm$~\,27 & 126$\pm$~\,5 & 293$\pm$~\,27 \\
$\tpz$           & 1278 & 2.83 & 384$\pm$102 & 233$\pm$18 & 151$\pm$104 \\
$\tpi$           & 3137 & 1.12 & 257$\pm$~\,24 & 5$\pm$~\,4 & 252$\pm$~\,24 \\
$\ppg$           &  770 & 0.91 & 377$\pm$~\,44 & 38$\pm$~\,6 & 339$\pm$~\,44 \\
\hline
\hline
\end{tabular}
\end{center}
\end{table}

\begin{table}[t]
\setlength{\tabcolsep}{0.3pc}
\catcode`?=\active \def?{\kern\digitwidth}
\caption{ For each $\eta$ channel, 
systematic uncertainties
in $\meta$ (in keV) from the listed sources (see text);
where applicable the degree of variation of the source
level is given (``Var'').
The sources marked with an asterisk ($^*$)
are assumed to be fully correlated across all modes;
others are assumed to be uncorrelated. 
The final column combines the uncertainties
across all modes with the weights given in the text. 
}
\label{tab:tablesys}
\begin{center}
\begin{tabular}{lrrrrrr}
\hline
\hline
Source & Var\ \  & $\gga$ & $\tpz$ & $\tpi$ & $\ppg$ & All\\
\hline
Fit Window             &                   & 14 & 52 & 12 & 22 &  9 \\
$M_\psip$$^*$          & 34~keV            & 27 & 32 & 25 & 32 & 27 \\
$M_\jpsi$$^*$          & 11~keV            &  9 & 16 &  9 & 10 & 9  \\
Bias                   & $\bias_i/3$       & 42 & 78 &  2 & 13 &  9 \\
Backgrounds            &                   & 19 & 67 & 17 & 27 & 12 \\
$p_{\pi^\pm}$ scale    & 0.01\%            &  1 &  4 &  5 &  1 &  3 \\
$E_\gamma$ scale       & 0.6\%             & 13 & 26 &  3 &  7 &  3 \\
MC Modeling$^*$        &                   & 46 & 46 & 46 & 46 & 46 \\
\ \ \ Sum              &                   & 74 &132 & 57 & 68 & 57 \\
\hline
\hline
\end{tabular}
\end{center}
\end{table}

\begin{thebibliography}{99}

\bibitem{rosner} J.L.~Rosner, Phys. Rev. D {\bf 27}, 1101 (1983).

\bibitem{escrib} R.~Escribano and J.~Nadal, J. High Energy Phys. JHEP05, 006 (2007).

\bibitem{michael} C.~Michael, Phys. Scripta {\bf T99}, 7 (2002). 

\bibitem{gerard} J.-M.~Gerard and E.~Kou, Phys. Lett. {\bf B616}, 85 (2005).

\bibitem{kekez} D.~Kekez and D.~Klabucar, Phys. Rev. D {\bf 73}, 036002 (2006).

\bibitem{PDG2006} W.-M.Yao \etal\ (Particle Data Group), J. Phys. {\bf G33}, 1 (2006).

\bibitem{PDG2002} K.~Hagiwara \etal\ (Particle Data Group), Phys. Rev. D {\bf 66}, 010001 (2002).

\bibitem{PDG1998} C.~Caso \etal\ (Particle Data Group), Eur. Jour. Phys. {\bf C3}, 1 (1998).

\bibitem{NA48} A.~Lai \etal\ (NA48 Collaboration), Phys. Lett. {\bf B533}, 196 (2002).

\bibitem{GEM} M.~Abdel-Bary \etal\ (GEM Collaboration), Phys. Lett. {\bf B619}, 281 (2005).

\bibitem{CLEO} 
  Y.~Kubota {\it et al.}  (CLEO Collaboration),
  Nucl.\ Instrum.\ Meth.\ A {\bf 320}, 66 (1992);
  M.~Artuso {\it et al.},
  Nucl.\ Instrum.\ Meth.\ A {\bf 554}, 147 (2005);
  D.~Peterson {\it et al.},
  Nucl.\ Instrum.\ Meth.\ A {\bf 478}, 142 (2002);
  CLEO-c/CESR-c Taskforces \& CLEO-c Collaboration, 
  Cornell University LEPP Report No.
  CLNS 01/1742 (2001), unpublished.

\bibitem{breta}
A.~Lopez \etal\ (CLEO Collaboration), Phys. Rev. Lett. {\bf 99}, 122001 (2007).
\end{thebibliography}
\end{document}